# Ion mobility and segregation in seed surfaces subjected to cold plasma treatments

*Alvaro Perea-Brenes,[a] Natalia Ruiz-Pino,[b] Francisco Yubero,[a] Jose Luis Garcia,[c] Agustín R. Gonzalez-Elipe,[a] Ana Gomez-Ramirez,[a,b] Antonio Prados,[b*] Carmen Lopez-Santos[a,d*]*

[a] Nanotechnology on Surfaces and Plasma Laboratory, Institute of Materials Science of Seville, Consejo Superior de Investigaciones Científicas-Universidad de Sevilla, Seville 41092, Spain

[b] Física Teórica, Departamento de Física Atómica, Molecular y Nuclear, Universidad de Sevilla, Apartado de Correos 1065, Seville 41080, Spain

[c] Department of Plant Biotechnology, Institute of Natural Resources and Agrobiology of Seville, Consejo Superior de Investigaciones Científicas, Seville 41012, Spain

[d] Departamento de Física Aplicada I, Escuela Politécnica Superior, Universidad de Sevilla, Seville 41011, Spain

*mclopez@icmse.csic.es; prados@us.es

ABSTRACT

Plasma treatment of seeds is an efficient procedure to accelerate germination, to improve initial stages of plant growth and for protection against pathogens infection. Most studies relate these beneficial effects with biochemical modifications affecting the metabolism and genetical growth factors of seeds and young plants. Using barley seeds, in this work we investigate the redistribution of ions in the seed surface upon their treatment with cold air plasmas. In addition, we investigate the effect of plasma in the lixiviation of ions through the seeds hull when they are immersed in water. Ions re-distribution in the outer layers of air plasma treated seeds has been experimentally determined through X-ray Photoelectron Spectroscopy analysis in combination with chemical in-depth profiling with gas cluster ion beams. The results show that in the shallowest layers of the seed hull (at least up to a depth of ~100 nm) there is an enrichment of $K^+$ and $Ca^{2+}$ ions, in addition to changes in the O/C and N/C atomic ratios. These data have been confirmed with the electron microscopy/fluorescence analysis of seed cuts. Observations have been accounted for by a Monte Carlo model simulating the electrostatic interactions that develop between the negative charge accumulated at the seed surface due to the interaction with the plasma sheath and the positive ions existing in the interior of seeds. Furthermore, it is shown that upon water immersion of plasma treated seeds mobilized ions tend to lixiviate more efficiently than in pristine seeds. The detection of a significant concentration of $NO_3^-$ anions in the water has been attributed to a secondary reaction of nitrogen species incorporated into the seeds during plasma exposure with reactive oxygen species formed on their surface during this treatment. The implications of these findings for the improvement of germination capacity of seeds are discussed.

# 1. Introduction

Plasma treatment of seeds has emerged as a common procedure to improve the seedling process due to its verified contribution to increase the germination rate, reduce the contamination degree by fungi or favor the initial stages of plant growth (1-3) through the interaction with soil microbiome (4), beneficial effects that are also noticeable under harsh environmental conditions (5, 6). To account for these effects, many studies have addressed the modifications induced in the biochemistry of seeds due to the plasma treatments (7-10) and the effect of reactive oxygen (ROS) and nitrogen (RNS) species in triggering biochemical metabolic processes that appear to modify the concentration of growth factors (11, 12), favor the generation of enzymes (13, 14), or even induce new gene expressions (15, 16). However, despite the numerous papers published during the last few years on this subject, the evidence is sometimes dispersed and focused on rather specific aspects. In general, it would be desirable a more holistic approach considering aspects such as biological variability, differences in experimental conditions between laboratories, or the influence of physical processes that occur during the plasma treatment of inorganic and polymeric materials (17-19). In relation with the latter, there is a clear gap of information regarding the physics that develops when a plasma discharge interacts with a biological system where mobility of ions can be affected by external electrostatic interactions, and similar effects. In this context, in previous studies on quinoa (20), cotton (21) or barley (22) seeds we have reported that $K^+$ ions may segregate towards the outer most surface layer of plasma treated seeds, and that similar diffusion process may affect other ions (23). Whether such a segregation is beneficial for the seedling of plasma treated seeds is still unknown. However, the evidence that the addition of moderate amounts of hexogen potassium to plants provides favorable conditions for the nitrogen metabolism, the photosynthesis and, in general, nutrient assimilation and transport, (24-28) suggests that plasma induced ion redistribution in the interior of seeds might also play an important role for the control of specific seedling processes.

This work primarily aims at determining the effect of cold atmospheric pressure plasmas in the distribution of charged ions in the interior of barley seeds. Secondly, we also investigate how the plasma treatment may affect the diffusion and/or lixiviation of key ions when treated seeds are soaked in water. The found changes in the distribution of ions in a relatively thick shallow zone of the plasma treated seeds are interpreted from a physical perspective, i.e., as consequence of the electrostatic interactions generated when a plasma discharge contacts a solid surface. The classical theory of cold plasmas proposes that when a plasma discharge interacts with a solid surface, a boundary region around the surface, called plasma sheath, is formed as consequence of the higher mobility of the electron with respect to the positively charged species in the plasma (29, 30), and that this leads to the accumulation of negative charge on the surface of the solid.

The main assumption of the present work is that the negative charge accumulated at the top-most seed surface during the plasma treatment may affect the distribution of ions in the seeds, attracting positively charged species (e.g., K+, Na+, $Ca^{2+}$,…) to their surface and repelling negatively charged ones (e.g., $OH^-$, $PO_4^{-3}$…) from it. In a similar way than in previous attempts to model the formation of ion diffusion profiles in polymers induced by the presence of electric fields (31), we formulate a Monte Carlo model accounting for the migration of positively charged species towards the seed surface during plasma treatment, to reach a canonical distribution of minimal free energy in an equilibrium state. To experimentally prove this unbalanced distribution of charged species in the interior of seeds, we analyze the chemical in-depth profiles of the outer layers of treated seeds with X-ray Photoelectron Spectroscopy (XPS) combined with gas cluster ion beam sputtering. This experimental approach is known to provide in-depth concentration profiles, preserving the chemical state of analyzing elements despite of the etching process (32). In fact, the use of this technique is recommended to minimize alterations in chemical composition of soft and sensitive materials, otherwise experiencing ion induced reactions or element preferential sputtering (33). This analysis has been complemented with the scanning electron microscopy observation and the electron X-ray dispersion analysis of seed cuts (21, 22).

In the experiments, barley seeds have been subjected to a high pressure dielectric barrier plasma discharge treatment (6, 22, 34). We confirm that the germination rate of treated seeds becomes accelerated by the plasma treatment. This result complements the evidence of an unbalance distribution of ions in the interior of seeds, the analysis of the lixiviation of $K^+$ and $NO_3^-$, and other minority ions upon immersion of treated seeds in pure water. Interestingly, this study shows that surface reaction and diffusion through the seed hull may be affected by the plasma treatment, confirming its importance not only to modify the biochemistry of metabolism, but also to alter physical processes that may significantly affect germination.

## 2. Materials and Methods

### 2.1. Plasma treatment and handling of seeds

Barley seeds (Hordeum vulgare L.) supplied by Intermalta SA (Spain) were exposed to an atmospheric pressure air plasma treatment in a parallel plate dielectric barrier discharge reactor at room temperature. Quartz plates of 0.5 mm thickness and 10cm diameter were placed onto two parallel stainless steel electrodes of 8 cm diameter separated by a distance of 4.2 mm. The top electrode was the active one, while the bottom electrode was grounded. A high voltage sinusoidal signal was applied to the top electrode by a TREK (model PD05034) high voltage amplifier connected to a function generator (Stanford Research System, model DS345). A sinusoidal signal of 1 kHz frequency and 8.6 kV voltage amplitude was set for the experiments. For these operating conditions, current amplitude was 6.5 mA and the discharge power 5.3 W. Ambient air at a pressure of 700 mbar was fixed as plasma gas in order to get reproducible treatments (actual pressure might slightly vary depending on environmental conditions in the laboratory (see additional details in. ref (22)). A set of 35 seeds were placed on the bottom electrode (grounded electrode) and exposed to the air plasma for 3 min at room conditions. A rough estimation of basic plasma features like electron density or temperature was done applying the Bolsig + code and the electron transport equation (35, 36).

Plasma species were characterized by Optical Emission Spectroscopy (OES) using an optical fibre connected to a monochromator (Horiba Ltd., Jobin Yvon FHR640). A diffraction grid with 1200 lines $cm^{-1}$ and entrance and output slits of 1 mm were used to collect the spectra with a photomultiplier working in the wavelength range from 200 to 700 nm. Operating conditions were 1 nm of spectral resolution and 0.5 s integration time.

## 2.2. Germination essay

The methodology employed for the germination essays in Petri dishes closely mirrors aspects of the industrial malting process of barley seeds. Groups of 50 seeds (pristine seeds or plasma treated seeds) have been carefully placed on a double layered Whatman filter paper located in 9 cm diameter Petri dishes. 4 mL of Milli-Q water have been employed for watering each dish. Afterwards, the Petri dishes were transferred to a bacteriological incubator set (Selecta model PREBATEM 80L) kept at 20 ºC in darkness. Petri dishes were meticulously examined at intervals of 12 h looking for signs of seed germination (penetration of the radicle through the seed coat or emergence of the coleoptile). Germinated seeds were tallied and transferred to separate Petri dishes each 12 h intervals during a 72 h cycle.

## 2.3. X-ray Photoelectron Spectroscopy depth profiling by means of gas cluster ion gun sputtering

X-ray photoelectron spectroscopy (XPS) depth profiling analysis of seeds was carried out using a Thermo Scientific 'K-Alpha' apparatus at the UK's National EPSRC XPS Users' Service (NEXUS) at Newcastle University, UK. Control (pristine) and plasma treated seeds at our lab in Seville were packed in plastic bags filled with nitrogen and sent to the NEXUS laboratory for testing within 24 h after plasma treatment. Three seeds of each type (plasma treated and pristine) were stick onto the sample holder using a double-sided carbon tape. This holder was placed in a pre-chamber and pumped until pressure was low enough to enable the transfer to the analysis chamber. Monochromatized Al K$\alpha$ radiation was used working at 12 kV acceleration voltage and 6 mA current emission. Acquisition of survey and single core level spectra were performed with 150 eV and 40 eV of pass energy, respectively. The binding energy (BE) scale of the spectra has been referenced to the C 1s signal of C-H/C-C signal at 284.5 eV. A spot size of 400 μm was used for the measurements.

Elemental depth profiles were measured at the same locations where survey and zone scans of the original samples were recorded prior to ion bombardment. The targeted depth of the profiles roughly varied from 100 nm to 200 nm depending on etching process applied (see below). Two etching protocols were applied to obtain the depth profiles:

Protocol a): depth profiling with a single charged Gas Cluster Ion Beam (GCIB) of 300 Ar atoms with an energy of 8000 eV (i.e., ~27 eV/atom), etching time of 120 seconds and estimated etching rate of 6.7 nm/cycle. Analysis time was 14 h to reach a depth of 100 nm, including 15 etching cycles.

Protocol b): initial low current monoatomic $Ar^+$ sputtering with 4 keV energy for a short time (i.e. a few minutes) to remove the outmost layers of the surface material. Depth profiling with GCIB of 1000 Ar with 6000 eV (i.e., ~6 eV/atom) and estimated etching rate of 21.6 nm / min per cycle. Analysis time was 7 h to reach a depth of 216 nm, including 10 etching cycles.

In both cases, survey and O 1s, C 1s, N 1s, Si 2p, P 2p, K 2p and Ca 2p spectra were recorded along these depth profiling experiments. Etching rates were calibrated with a $Ta_2O_5$ standard film of known thickness.

The fitting analysis of the C 1s and O 1s high resolution spectra was carried out under the assumption of a series of chemical species identified with the following binding energies: C-C (284.5 eV), C-OH (285.5 eV), C-O-C (286.5 eV), C=O (287.5 eV), and COO (288.4 eV) for the carbon region and C-O (532.7 eV), C=O (533.2 eV), N-O (536.2 eV) for the oxygen region (37, 38). Since the oxygen of Si-O bonds contributes to the O 1s signal, this oxygen was subtracted from the overall oxygen content to estimate the amount of oxygen associated exclusively to water or other organic components. An estimation of this oxygen was made taking the Si 2p signal as a reference and a $SiO_2$ stoichiometry for the Si-O compounds (note that this is a simplification because other C-Si-O bond structures can be present in the examined samples).

**2.4. Scanning Electron Microscopy and Energy Dispersive Spectroscopy analysis**

Scanning Electron Microscopy (SEM) and Energy Dispersive Spectroscopy (EDX) analyses have been done for the surface and seed cuts (i.e., cross section) of untreated and plasma treated barley seeds. A Hitachi S4800 SEM-FEG field emission microscope working at 2 kV has been used for morphological characterization, without applying any metallization protocol. Elemental EDX maps have been obtained in an EDX-Bruker-X Flash-4010 analyzer working at 20 kV. Analyses have been done for the surface and the seed cuts (i.e., cross section) of the barley seeds just after the plasma treatment in comparison with the untreated seeds. Seed cuts were prepared by carefully slicing the seeds with a blade and are analysing in the microscope within a time interval shorter than 30 min after plasma treatment.

**2.5. Ion release by soaking water. Analysis by Inductively Coupled Plasma Mass Spectrometry**

To study cations (potassium, sodium, calcium, magnesium, and zinc) and anions (nitrates, chlorides, and phosphates) release in water from the plasma treated seeds during water immersion, we have used the Agilent 7800 system from the ICP-MS Analysis Service of the Institute of Natural Resources and Agrobiology of Seville (IRNAS-CSIC). For the analysis of anions and ammonium, a segmented flow auto-analyzer, model BRAN-LUEBBE, was used during the ICP-MS test. The study was done for original seeds without any treatment and for seeds treated with plasma. A number of hundred seeds with weights ranging between 4-5 g were used for each experiment. Milli-Q water was used as soaking medium. The volume used for each experiment was slightly corrected considering the weight of each group of 100 seeds, being in the order of 40-45 ml depending on experiment. To check the release of ionic species, the seeds were immersed in the milli-Q water for 10 and 120 min. Then, the seeds were extracted and an aliquot of 10 ml of the liquid was taken for ICP-MS analysis. No more than 1 hour elapsed since the recovery of the aliquots and their analysis by ICP-MS. To ensure a safe and uncontaminated delivery, the aliquots were stored and handled in closed tubes under dark conditions. Results are referred to the weight of seeds and will be expressed in mg/g (i.e. weight of detected species per gram of seeds).

**2.6. Simulation model of ion migration in plasma treated seeds**

The diffusion of positive ions in the seed and its hull and the resulting steady-state distribution of positive ions after the plasma treatment were simulated using a simplified Monte Carlo algorithm. A basic assumption of the developed model is that the seed surface becomes negatively charged as the result of the formation of a plasma sheath during exposure to the plasma (39). The aim of the model is to simulate the effect of the electrical field associated to this negative charge accumulated on the surface on the positively charged species (i.e., cations) distributed in the interior of seeds and thus account for the new physico-chemical processes that may take place during plasma-surface interaction. It is a basic assumption of the model that at least part of the negative charge coming from the plasma will remain on the surface and will not diffuse or recombine with positive charges already present in the system.

The model relies on a two-dimensional (2D) circular representation of the seed and on the segmentation into radially distributed cells of equal area. Positive ions are treated as point charges that are able to move throughout the seed. Before the treatment, the seed is electrically neutral and therefore the $N$ positive ions with charge $q$ are assumed to be randomly distributed inside the circular area simulating the seed. A negative density of charge, $\sigma$, compensates the charge if positive species. It is assumed that this density of charge is uniformly distributed all over the "seed" disk. A schematic of the model is presented in the Supporting Information Figure S1. The following ratio should hold to account for the electrically neutral character of the pristine seeds in its initial state:

$$Nq + \sigma \pi R^2 = 0 \quad (1)$$

where $R$ is the radius of the circle representing the seed. For the calculations, the radius has been taken equal to 1 mm. Since we approximate the real problem to a two-dimensional model of the seed, its "surface" would be the perimeter of the circle, i.e., a circumference of radius $R$ that bounds the interior of the seed. To simulate the effect of the plasma treatment and the accumulation of negative charge on the seed surface during plasma treatment, randomly distributed locations have been chosen to accumulate the extra negative charge associated to the formation of the plasma sheath. More specifically, we assume the accumulation of $N_{ext}$ negative charges with charge $Q_{ext} < 0$ distributed at random positions $\vec{R}_j, j = 1, \dots, N_{ext}$, on the surface of the seed, i.e.,

$$\vec{R}_j = R \left( \cos \theta_j \ \vec{u}_x + \sin \theta_j \ \vec{u}_y \right) \quad (2)$$

with $\theta_j$ being a random number uniformly distributed in $[0, 2\pi]$. This "extra" negative charge, created by the plasma treatment, is assumed to be fixed at the initial positions and thus they do not move during the Monte Carlo simulation (Figure S1b in the Supporting Information).

The Monte Carlo algorithm is then run looking for the minimization of Helmholtz's free energy $F$ of the system, once the negative charges have accumulated on the surface during the plasma treatment. We recall that $F = U - TS$, where $U$ is the internal energy, $T$ is the temperature, and $S$ is the entropy. The internal energy $U$ comprises several terms, stemming from all possible electrostatic interactions of the mobile positive ions in the interior of seeds, the background negative charge homogeneously distributed in the interior of seeds, and the fixed negative punctual charges accumulated at the surface. These immobile negative punctual charges will interact electrically among themselves and with the negative charge background density in the interior of seeds. However, from the point of view of the Monte Carlo calculations, the energy associated with this interaction is constant and does not evolve from an iteration to the next. In other words, this negative-negative charge interaction can be ignored for the simulations (the Monte Carlo scheme only considers energy differences between consecutive configurations). Therefore, calculations of the total energy are restricted to the electrostatic interaction (i) between the negative charges at the surface and the mobile $N$ positive charges inside the seed, (ii) between the mobile positive charges themselves, and (iii) between the mobile positive charges and the background negative charge, as detailed below. The sum of these interactions can be formulated as:

$$U(\vec{r_1}, \dots, \vec{r_N}) = \sum_{i=1}^{N} U^{(i)} \quad (3)$$

where $U^{(i)}$ is the contribution of the $i$-th positive ion, characterized by its position $\vec{r_i}$, to the internal energy. More specifically, $U^{(i)}$ can be split into three terms: (i) $U_1^{(i)}$, which accounts for the interaction with the fixed negative points of charge $Q_{ext} < 0$ on the surface boundary of the seed, (ii) $U_2^{(i)}$, which accounts for the interaction with the other mobile positive ions of positive charge $q$ in the interior of the seed, and (iii) $U_3^{(i)}$, which accounts for the interaction of positive ions with the negative background charge upon the redistribution of the former:

$$U^{(i)} = U_1^{(i)} + U_2^{(i)} + U_3^{(i)} \quad (4)$$

$$U_1^{(i)} = k \, Q_{ext} \, q \sum_{j=1 \dots N_{ext}} \frac{1}{|\vec{r_i} - \vec{R_j}|} \quad (5)$$

$$U_2^{(i)} = \frac{k q^2}{2} \sum_{\substack{j=1...N \\ j \neq i}} \frac{1}{|\vec{r_i}-\vec{r_j}|} \qquad (6)$$

$$U_3^{(i)} = q V(\vec{r_i}) \qquad (7)$$

where $k = (4\pi\epsilon_0)^{-1}$ and $V(\vec{r})$ is the potential created by the background charge density $\sigma$, in the interior of the "seed" disk. The resulting curve of this potential is presented in the Supporting Information Figure S2.

Since the experiments are done at constant (room) temperature $T$, we assume that the seeds reach the equilibrium state corresponding to this temperature. At equilibrium, the probability distribution has the canonical form, i.e. the probability density of finding the positive ions in the configuration $\{\vec{r_1}, \vec{r_2}, ..., \vec{r_N}\}$ is given by the canonical distribution

$$\rho(\vec{r_1}, \vec{r_2}, ..., \vec{r_N}) = \frac{1}{Z} e^{-\beta U(\vec{r_1}, \vec{r_2}, ..., \vec{r_N})} \qquad (8)$$

where $\beta = (k_B T)^{-1}$, with $k_B$ being Boltzmann's constant, $T$ the absolute temperature (i.e., in Kelvin), and $Z$ is the partition function, which ensures the normalization of the probability density,

$$Z = \int d\vec{r_1} \int d\vec{r_2} ... \int d\vec{r_N} \, e^{-\beta U(\vec{r_1}, \vec{r_2}, ..., \vec{r_N})} \qquad (9)$$

We recall $F = - k_B \ln Z$. In order to reach the equilibrium state, we introduce an effective stochastic dynamics that drives the system to equilibrium. Details about this procedure are provided in the Supporting Information S3. In the simulation, we assume that there exists an energy barrier $E_0$ for the positive ions moving between neighboring cells, i.e., when crossing the cells boundary, a process that would have an energetic cost in a real situation. This activation energy thus accounts for the diffusion constraints in the seeds that hinder the movement of the positive ions. After some relaxation time, the *N* positive charges reach a final equilibrium distribution, which does not evolve (in average, there are thermal fluctuations). In an actual seed, this "final" cation distribution state resulting from the plasma treatment might be further disturbed when exposing the seeds to new conditions (e.g., immersion in water, exposure to humid environments or high temperatures, etc.).

## 3. Results and Discussion

### 3.1. Air DBD plasma properties and optical characterization

A rough estimation of the electron flux arriving at the seed surface was done using the Bolsig+ code, and the electron transport equation. For this basic analysis we disregarded possible collisional ionization and/or recombination processes that might take place within the plasma sheath at the experimental atmospheric pressure conditions of the experiment. Values of 3.74 eV, $4.8 \cdot 10^{13}$ m$^3$ and $8.28 \times 10^{-2}$ m$^2$V$^{-1}$s$^{-1}$ were estimated for the mean electron temperature, electron density and electron mobility, respectively. These data permits to calculate the flux of impinging electrons on the electrode surface and the seeds placed thereon. The obtained value was $2.87 \times 10^{16}$ electrons m$^{-2}$s$^{-1}$ under our experimental conditions.

The air plasma used for seeds activation was also characterized by OES to identify some of the active species that, generated in the plasma phase, will chemically interact with the seeds. Figure 1a shows a typical OES spectrum recorded when treating the barley seeds with atmospheric air plasma. In Figure 1b we show a picture of the interior of the reactor with the seeds exposed to the plasma. The spectrum is dominated by the second positive system of the excited neutral molecule $N_2^*$ [$C^3\Pi \rightarrow B^3\Pi$] in the range of 290–385 nm (40). Notably, these species do not have an oxidative character, although may interact chemically with the surface of the hull. It is important to stress that besides these species detected by OES, it is most likely that other chemically active species will also form in a gas mixture of $N_2$, $O_2$ and other minority components. According to other works in the field (41, 42), it is expected that nitrogen oxide molecules (NO$_\gamma$), typically appearing in the region 225–283 nm, OH* species or O$_x$* species appearing around 287-309 nm and 715-780 nm respectively also form in the air plasma used to activate the seeds. In our case, these species have not been efficiently detected either because their emission lines do not have enough intensity (note that seeds inside the plasma discharge produce a decrease of the light intensity collected by the fiber) or because they are so short lived that their emission spectrum results undetectable. We should mention in this regard that in a previous work on barley seeds we showed that ROS species resulting from the interaction with plasma species of oxygen accumulate in the seeds as result of the plasma treatment. The chemical effect of these ROS could be emulated by hydrogen peroxide, suggesting that they are diatomic species of oxygen, likely peroxide or superoxide, both with a significantly high oxidative character (22).

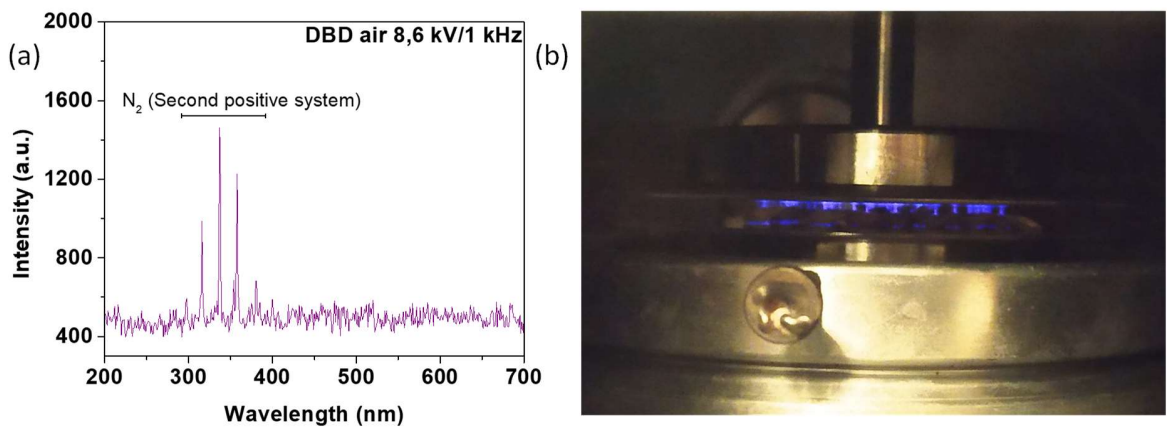

Figure 1. (a) Optical emission spectra corresponding to the air plasma discharge in the presence of barley seeds. (b) Picture of the air plasma treatment on the barley seeds.

### 3.2. Germination rate of plasma treated barley seeds

The plasma treated barley seeds behave in a similar way to those in previous experiments carried out with this type of seed (6, 22). In agreement with the already reported results, plasma treated seeds experienced an acceleration in the germination rate as illustrated in Figure 2. This acceleration is particularly noticeable after the first 18 h, where the number of germination events was 25% higher for the plasma treated with respect to pristine control seeds. In previous works, we have verified that in addition to this increase in germination rate there were other changes in the phenotype of young plants when the seeds were sawn in soil. These changes, both in germination rate and in seedling process have been widely studied and commonly attributed to the chemical interaction of ROS (reactive oxygen species) and RONS (reactive oxygen and nitrogen species) accumulated in the seeds during the plasma-seed interaction with the normal metabolic chains of seeds and young plants (43, 44). Unlike this common approach, herein we focus the attention on the changes in the distribution of ions in the interior of seeds. This analysis has been carried out by XPS complemented with the SEM-EDX analysis of cross section cuts of the seeds.

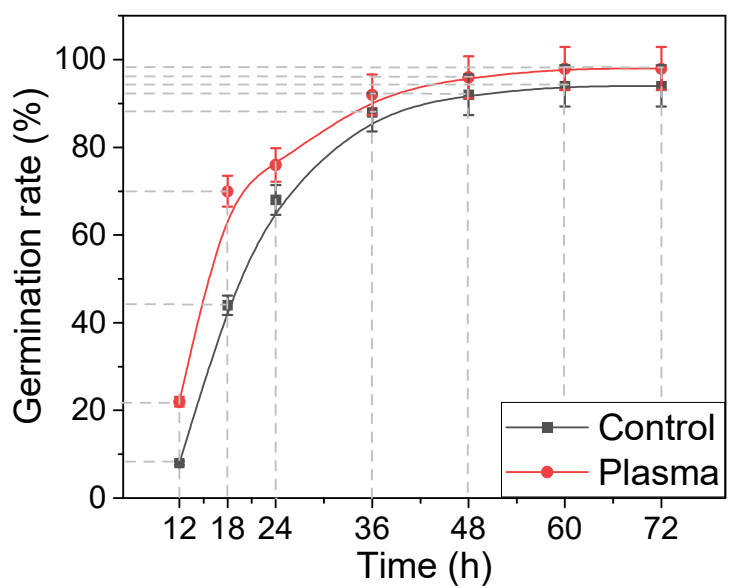

Figure 2. Germination rate in percentages for pristine (control) and plasma treated barley seeds as a function of time since they were put in the Petri dishes.

3.3. In-depth chemical analysis of plasma treated seeds

The plasma treatment herein applied is similar to that used in previous studies on barley seeds, and as expected, it leads to similar beneficial effects on seed germination and plant growth (6, 22). In this work, we have specifically investigated the changes in surface composition that are produced by the plasma treatments. The gentle plasma conditions used to treat the seeds did not induce damage signs on the seeds surface morphology, as can be appreciated in the SEM/EDX analysis reported in Figure 3 and in the Supporting Information S4. It is noticeable that the SEM images of seed cuts included in these two set of figures reproduce well the main morphological features of the inner structure of Barley seeds previously reported by Geng et al and that this internal morphology is not affected by the plasma treatment (45) However, comparison of the EDX maps of the seed cuts in Figure 3 taken before and after plasma treatment show a certain hull enrichment of elements such as K (see the more intense color of hull zone in the K distribution zoomed map after plasma treatment). The thickness of this enrichment zone could be estimated to be of several tens of microns (~ 40 µm (46)). By this analysis, it is also important to highlight that SEM/EDX reveals the presence of silicon nodules at the surface of barley seeds (22). This observation will be taken into consideration when analyzing the O 1s photoelectron spectra recorded during the depth-profiling analysis by XPS. Additional elements like Mg, Na, Cl, Al, Ca, and S have been detected on both treated and untreated seed cuts (see supporting information, Figure S5) although without clear evidence of accumulation or depletion at the hull region. It is important to remark that the compositional distributions presented in Figure 3 did not significantly vary upon storage in ambient conditions for some days after the plasma treatment. This supports the idea that the accumulation of potassium on the external region of the seed hull must proceed through diffusion from inner regions of the seed.

XPS provides a more precise analysis of the chemical composition of most external regions of seeds surface. Table 1 summarizes the element concentration in the outmost outer layers monitored by this technique for the pristine and plasma treated seeds before depth profiling. The probe depth with a conventional XPS analysis can be roughly estimated in 1-2 nm thickness (i.e., a very thin surface region). It is apparent in this table that the surface composition of this outer layer drastically changes after plasma treatment. Similar changes in XPS compositions have been previously reported for barley, cotton and quinoa seeds subjected to plasma treatments (20-22).

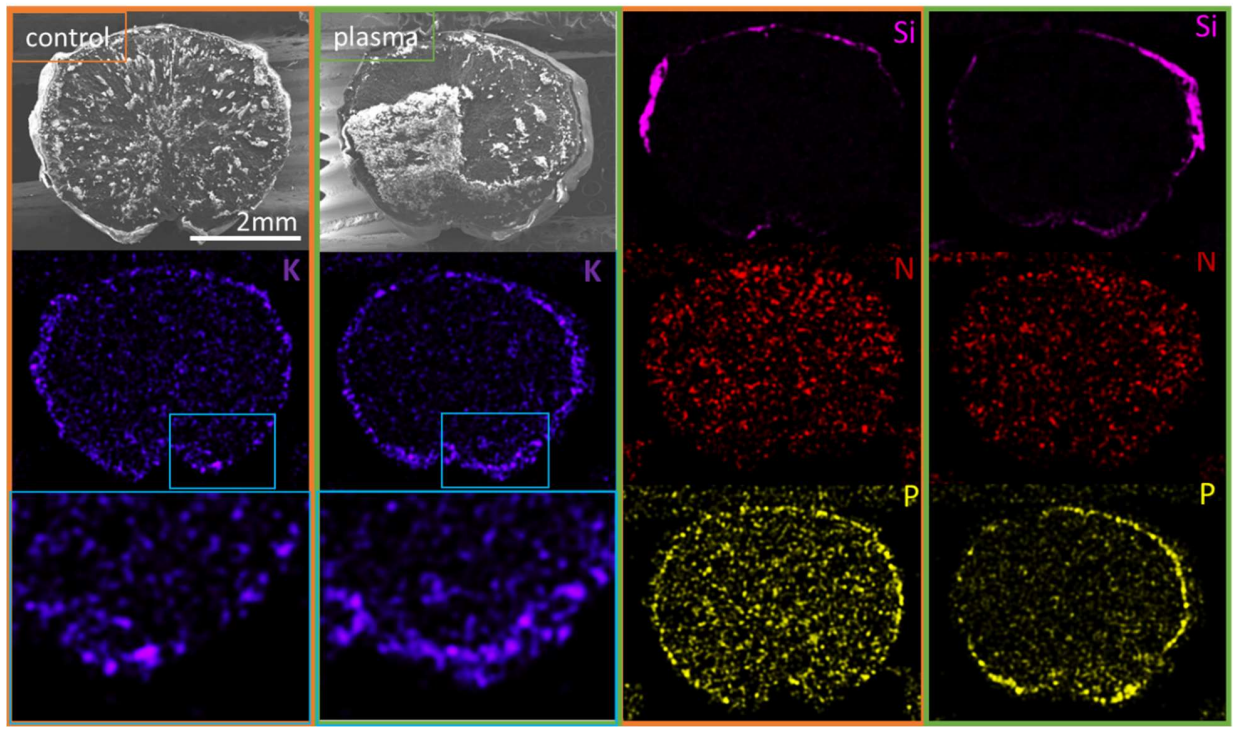

Figure 3. SEM-EDX analysis (K, Si, N, P maps, including a couple of enlarged zones of the K maps) of cross sectional cuts of the seeds showing the enrichment of some minority elements in the outer zones of hull after the plasma treatment: (a) control barley seed; (b) barley seed exposed to air plasma treatment for 3 min.

The common tendencies observed when comparing atomic concentrations can be summarized as follows: plasma treatment induces a decrease in the carbon and an increase in oxygen and, for barley surface, also in silicon. It is also noticeable the increase in the percentages of minority elements such as K, Ca and N after plasma treatment. The increase in oxygen and the decrease in carbon are a direct effect of the interaction of carbon containing materials with oxygen containing plasmas and, are the result of the partial oxidation of organic matter and the formation of -C=O, -C-OH or -COOH species (47, 48). It is also likely that complete oxidation may also affect the organic matter in the outermost atomic layers of seeds leading the formation of $CO_2$ and $H_2O$.

Such a series of effects have been widely reported for polymer materials and, as mentioned before, also for seeds. At the surface of barley seeds, Figure 3 and Figure S4 in the Supporting Information show the presence of a relatively large number of spotted $SiO_2$ agglomerates, rendering a significant average concentration for Si at the surface, as monitored here by XPS. The increase in the Si concentration after plasma treatment is likely the result of the cleaning removal of adsorbed airborne hydrocarbons or other carbon contamination on the surface of seeds, as well as the possible removal of an organic layer covering these agglomerates. We assume that these $SiO_2$ agglomerates are inert and do not play any significant role in the ion redistribution processes that are the subject of the present study. Therefore, regarding plasma reactivity and the possible effect in seed germination, the most significant feature is the increase in K, Ca and N concentrations found for the plasma treated seeds.

Table 1. Percentage of elements determined by XPS in the pristine and plasma treated seeds

|   | Atomic percentage (%) | |
|---|---|---|
|   | Control seed | Plasma treated seed |
| C | 88.5 | 52.8 |
| O | 9.2 | 37.6 |
| Si | 0.4 | 4.8 |
| N | 0.8 | 2.1 |
| K | 0.6 | 1.7 |
| P | 0.2 | 0.2 |
| Ca | 0.3 | 0.8 |

Since conventional XPS only provides information over the outermost surface layer of samples (i.e., approximately 1-2 nm of thickness) it is important to determine whether these or similar changes also affect the inner layer of the seeds. With this purpose, a series of depth profiling experiments have been carried out using the two experimental protocols described in the experimental section. Cluster ion depth profiling according to *protocol 1* permits monitoring the evolution of elements concentration along a layer of ca. 100 nm thickness. Figure 4 shows a series of depth profiles recorded for a reference and a plasma treated seed. Depth profiles are plotted as a function of GCIB etching time. Data are reported for the majority elements C, O and Si in Figure 4a and b. The depth profile recorded for the control seed reveals an enrichment in carbon in the outermost layer of seeds (some contribution by air borne carbon contamination cannot be discarded), and a progressive and smooth enrichment in O together with a relative depletion in C at deeper zones of the seed. However, since the Si concentration also increases with depth, part of this increase in oxygen should be associated to the $SiO_2$ agglomerates present at the seed surface. In practice, discounting the percentages of silicon and associated O, it appears that carbon concentration is rather constant all along the etched thickness, with average values around 55%. This evolution of C concentration agrees with the known structure and the thickness of several microns of seed hull, which is known to be mainly formed by cellulose, tannins and small amount of some polyphenols, i.e., compounds very rich in carbon (49).

This situation drastically changed for the seed exposed to the plasma. The depth profiles in Figure 4b indicate that, once discounting the oxygen associated to silicon, there is still a non- negligible amount of oxygen with values up to 11.5 % that distributes all along the depth of the etching profile. This renders an O/C ratio varying from 0.13 at the surface to 0.28 after etching 100 nm of thickness. In other words, a considerable amount of oxygen has become incorporated into the outer layers of seed hull, extending along more than 100 nm. The etching profiles obtained using the protocol 2 (see Supporting Information S6), confirm that oxygen becomes incorporated in the carbon rich matrix of the seed hull and that this enrichment in oxygen occurs at least up to a thickness of 216 nm. The oxygen incorporation into organic materials upon plasma exposure is a common feature for polymers and related materials (50, 51).

However, the most interesting issue in relation to the main cue of this work, i.e., that charged ions may migrate towards the surface upon plasma exposure, is the analysis of the minority elements also recorded during the depth profiling analysis. Depth profiles of minority elements K, Na, Ca, N and P are reported in Figures 4c and d for the original and plasma treated seed, respectively. Most significantly, the concentrations of K, Ca and N along the whole depth profile are higher in the plasma treated than in the control seed, indicating a net enrichment in these elements after the plasma treatment. Moreover, the profiles for K and Ca differ in shape with respect to that in the control seed where the concentration profiles vary smoothly. We propose that the detected enrichment in K and Ca in the outermost layers of the seed hull (i.e., at least up to 100 nm in this experiment) results from electrostatic interactions developed between the negative charges accumulated on the surface of the seeds during the plasma treatment and the relatively free and mobile positive ions ($K^+$, $Ca^{2+}$, $Na^+$) existing in the interior of seeds. Besides Ca and K, another minority element whose distribution increased in this experiment along the examined thickness of seeds was nitrogen. Meanwhile, P distribution remained almost unaltered after the plasma treatment.

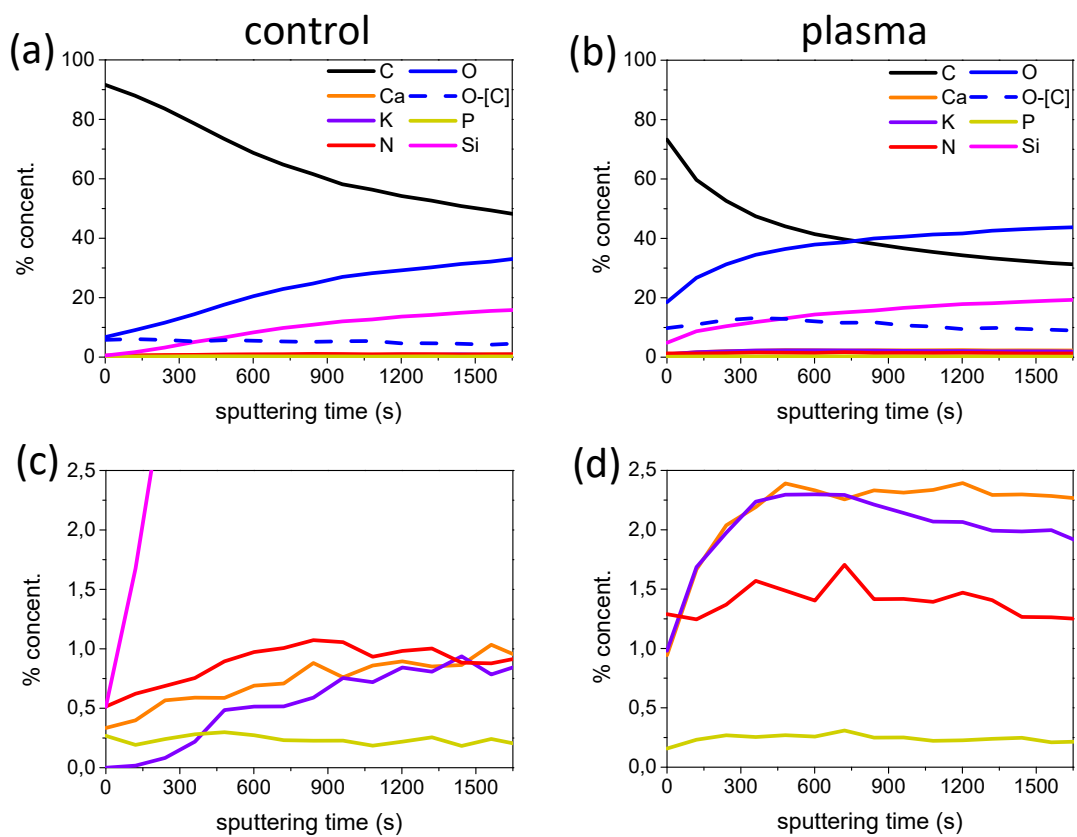

Figure 4. Element depth profiles for the control ((a), (c)) and plasma treated ((b), (d)) seeds (up to 100 nm in depth). Dashed lines represent the depth profile of oxygen associated to the organic component of the outer layers of seeds (i.e., once discounting the oxygen associated to silicon)

To properly explain the observed changes in the concentration of the detected elements at the outmost surface layers of the plasma treated seeds, we deem it necessary to get a deeper understanding of the chemical nature of the species detected by XPS. Figure 5 shows selected C 1s, O 1s and N 1s spectra taken along the depth profiles in Figure 4 both for the reference and plasma treated seeds. Equivalent series of spectra for the Si 2p, K 2p, Ca 2p and P 2p levels are reported as Supporting Information Figure S7. The binding energy values assigned to the main functional groups of the elements detected at the barley surface before and after exposure to the plasma discharge are gathered in Table S8 in the supporting information. .Particularly, silicon contribution detected at the surface of the plasma treated seed is located at 102.8 eV of binding energy, a value pointing up to the presence of Si-C-O groups. This result agrees with previous works quantifying the silicon content of different varieties of barley seeds. (52, 53) After etching, peak shifts to 103.4 eV, a BE value typical of $SiO_2$ (54). The appearance of doublet peaks around 292.8 eV and 295.6 eV (55) and 346.7 eV and 350.2eV (56) are an indication of the presence of $K^+$ and $Ca^{2+}$, respectively. The shape of the spectra/BE of $2p_{3/2}$ peaks do not change along the depth profile, supporting that the same chemical species appear distributed along the whole analyzed thickness of the plasma treated barley seed. Otherwise, the photopeak around 132.7eV is commonly assigned to the P-O functionalities, likely due to phosphate anions (57).

The series of spectra in Figure 5 indicate a progressive variation of the O 1s, C 1s and N 1s spectral shapes with the etching time. These changes occur in parallel to the quantitative variations in atomic percentages reported in Figure 4. To get more precise information of the chemical characteristics of the chemical bonds present in the examined layers of the seeds and how they vary with the etching time, a fitting analysis was carried out for selected spectra taken from Figure 5. Fitting of the C 1s spectra was carried out under the assumption of the contribution of the following functional groups: C-C (284.5 eV), C-OH (285.5 eV), C-O-C (286.5 eV), C=O (287.5 eV) and COO (288.4 eV) (37, 38, 58) (see Figure 6).

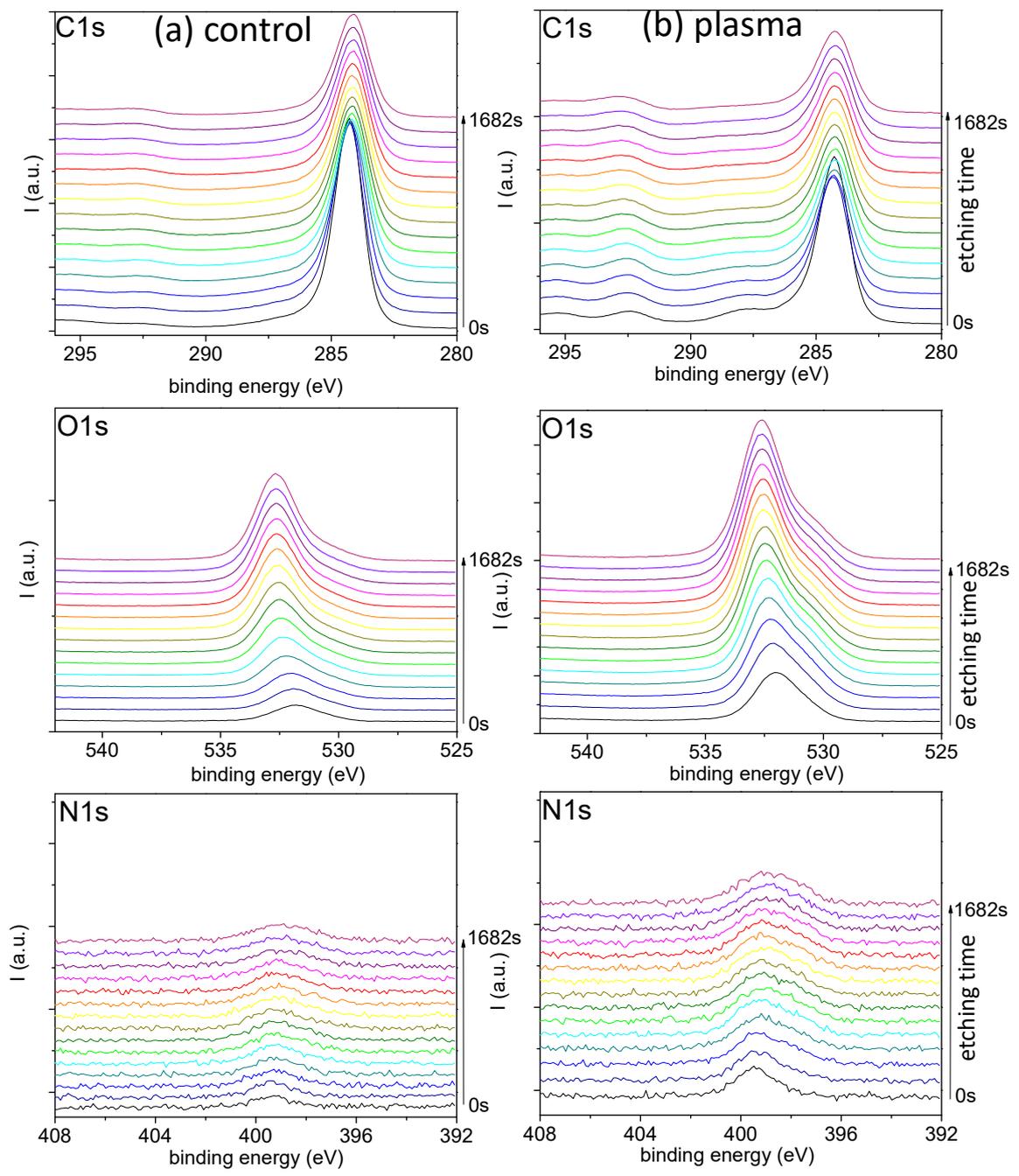

Figure 5. Selected C 1s, O 1s and N 1s spectra taken along the depth profiles (i.e., progressive etching times from 0 to 1682 seconds as indicated with the arrows) for control (a) and plasma treated (b) barley seeds.

We must note that the contribution of C-N functional groups will overlap in the region around 286.2 eV and has not been explicitly considered to simplify the fitting analysis because it would be indistinguishable from the C-O contribution (59). The C=O and COO functional groups were relatively more intense at the surface of the plasma treated seeds (see the enlarged representations of the spectra at 0 s etching time in Figure 6c and d). This can be taken as a hint of the plasma etching oxidation of the outer carbonaceous layers of the seeds. Significantly, Figure 6c also shows that the concentration of these groups decreases in the deeper layers of the plasma treated seeds that are examined after 1600 s of etching time. In addition, since these deeper zones of the seeds are not directly exposed to the plasma, we assume that the small relative increase in C=O and COO functional groups in the plasma treated seeds that can be deduced when comparing the 1600 s spectra in Figure 6c and d must proceed from a certain rearrangement mechanism of carbon functional groups and/or the diffusion of reactive species of oxygen from the surface towards the interior of seeds.

The O 1s spectra (c.f, Figure 6e), can be fitted with a series of bands that can be attributed to O-Si / O-C (at 532.8 eV), HO-C (532.0 eV) and O=C (530.8 eV) (60, 61). The intensity of the O 1s band attributed to O-Si bonds has been estimated from the measured intensity of the Si 2p peak. The most significant difference between control and plasma treated seeds is the increase in the intensity of the band at around 530 eV due to surface oxidation of carbon. Interestingly, the contribution at around 532.8eV, which are compatible with O-Si bonds, and contributions at higher BEs might be also due to some form of diatomic oxygen of peroxide or superoxide type (62). These chemical species are usually taken as expression of reactive oxygen species (ROS), which have demonstrated to be very active in triggering metabolic reactions beneficial for the germination and seedling of seeds (63).

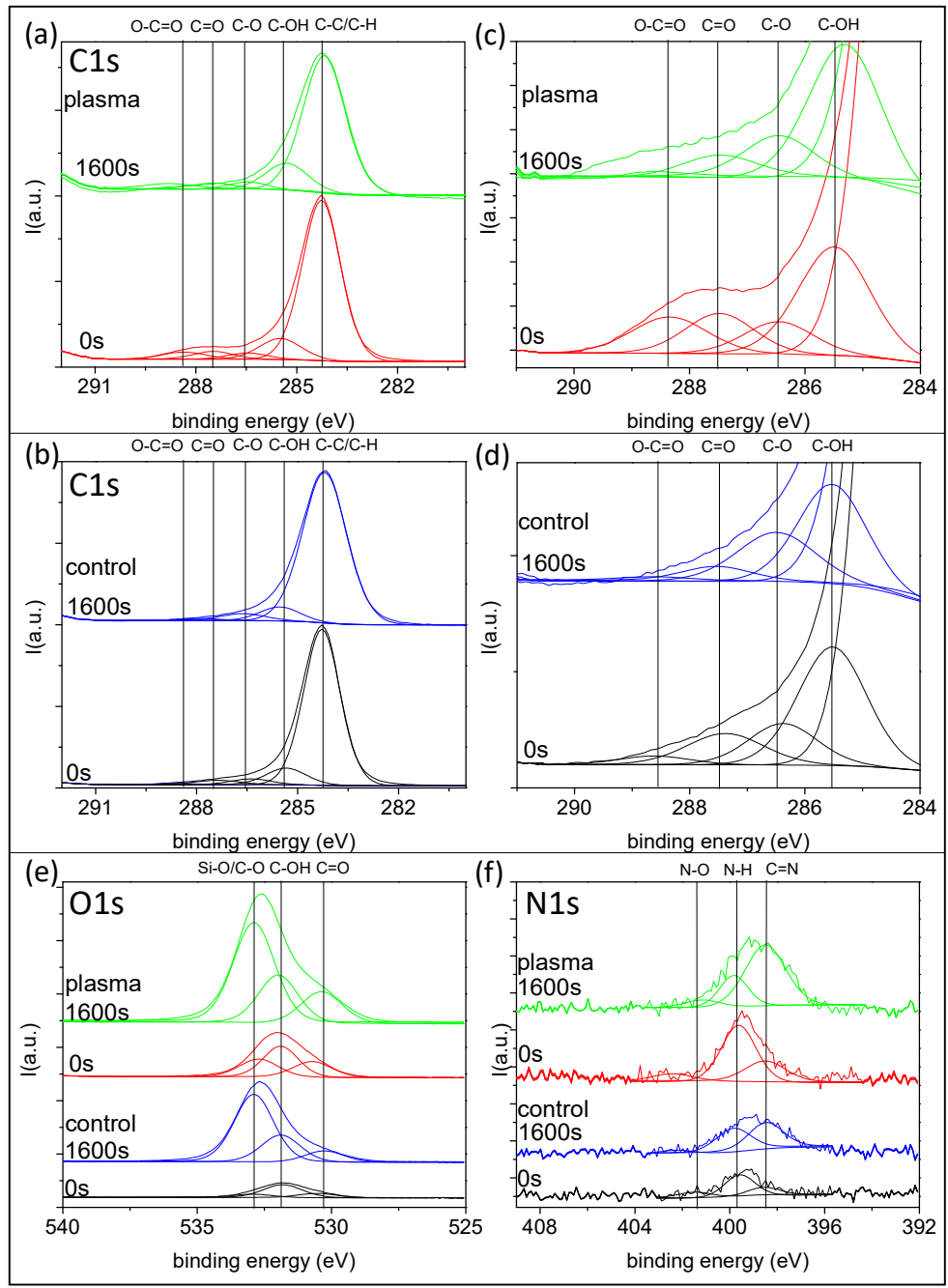

Figure 6. Fitting analysis of selected spectra taken from Figure 5 corresponding to the first (i.e., 0 s) and last (i.e. 1682 s) spectra of the series, both for the control and plasma treated seeds: (a), (b) C 1s regions ; (c), (d) zoomed C 1s regions; (e) O 1s regions; (f) N 1s regions. Spectra intensities have been normalized to the C 1s control 0s area.

The N 1s spectra (c.f., Figure 6f) have been fitted under the assumption of three bands attributed to N=C (398.5 eV), C-N-H$_x$ (amine at 399.2 eV) and protonated amine (C.N-H$_{x+1}^+$) or N-C=O (amide) groups (401.3 eV) (64). By this analysis, it is relevant to highlight that there are not clear contributions at BEs higher than 402 eV, a region where nitrogen-oxygen bonds appear in the N1s photoelectron spectra (NO$_x$ or similar species are characterized by BEs as high as 405 eV). Therefore, this analysis discards the existence of significant concentrations of RONS species on the surface of the seeds, even after the plasma treatment. We assume that the observed increase in nitrogen (c.f. Table 1 and Figure 4) is due to nitrogen species bonded to carbon very likely through direct interaction of the carbon rich structure of the hull with excited nitrogen species of plasma such those detected by Optical Emission Spectroscopy (see Figure 1). It is noteworthy that the found increase in nitrogen concentration extends through, at least, a thickness of 200 nm of seeds (see Supporting Information S6), thus indicating that the effect of plasma does not remain confined at the outmost surface layer but extends to the interior of seeds. Interestingly, this effect is less pronounced for the oxygen linked to carbon, whose concentration, within the incertitude limits imposed by the presence of SiO$_2$/Si-C-O agglomerates, only starts to decrease after the first 80 nm of thickness (see Figures 4 and S6).

### 3.4. Simulation of positive ion redistribution during plasma-seed interaction

The Monte Carlo model described in the Experimental section and Supporting Information S3 has been applied to account for the redistribution of positive charges between the inner and outer layers of seeds upon plasma activation. This simulation relies on the hypothesis that an extra negative charge accumulates on the surface of seeds during this process and that the developed electrostatic interactions are responsible for the observed surface enrichment in $K^+$ and $Ca^{2+}$ ions detected by XPS. The application of the Monte Carlo algorithm permits following the evolution in the distribution of positive charge from its initial homogenous state prior to plasma exposure to a final equilibrium state of minimum free energy. Typical configurations for the initial and final states of the system are shown in the schematics in Figure 7a. The developed Monte Carlo algorithm has been computed using the following parameter values:

$N = 10^3$ ions, $N_{ext} = 10$, $Q_{ext} = 4 \times 10^5 q$, $R=1$ mm, $T = 300$ K, and $E_0 = 4k_B T$.

These values do not reproduce the actual conditions of the experiment but aim at providing a reference frame of the physical processes occurring in the seed due to the plasma treatment. Neither the model size (i.e., R) nor the amount of accumulated charge (i.e. $N_{ext}$ x $Q_{ext}$) are in direct correlation with the actual size of a barley seed or the total amount of negative charge impinging on the seed surface during the plasma treatment time that, taking into account the estimated flux of electrons, would be in the order of $3 \times 10^{16}$ electrons m$^{-2}$. Thus, although the values selected for $N_{ext}$ and $Q_{ext}$ has been chosen arbitrarily, they comply with the condition that, in absolute values, $N_{ext} Q_{ext} \gg Nq$. In other words, the model assumes that during the plasma treatment the extra negative charge accumulated on the seed surface due to the higher mobility of electrons, compared to that of the positively charged species in the plasma, is much larger than the mobile positive ions existing in the interior of the seed. This condition ensures that electrostatic interactions developed during the plasma treatment are the main factor inducing the migration process of the cations towards the surface.

The value selected for the energy barrier corresponds to a potential $V_0 = E_0/q$ equal to 0.1034 V, i.e., a value within the same order of magnitude than the resting membrane potential in plants (65).

An additional remark should be made with respect to the $N_{ext}$ spots at the surface that concentrate all negative charge $Q_{ext}$. They figure out the distribution of charge deposited on the shell of the seed during the plasma treatment. In real seeds, the shell does not have a perfectly smooth morphology but has "cusps" at which the charge would be preferentially deposited because of the well-known sharp-point focusing effect on the electric field. This justifies why the charges $Q_{ext}$ accumulate in the model at $N_{ext}$ random points along the perimeter of the circle, as defined by Equation 2. In our simple model, to lower the computed number of interactions and to accelerate the convergence of the algorithm, we have considered a limited number of charged points (i.e., $N_{ext} = 10$) each one with a "high" accumulation of negative charge, i.e. $Q_{ext} \gg q$.

Once the system has reached its equilibrium state in the Monte Carlo simulations, the positive charges representing the cations in the seed appear to undergo a redistribution consisting of a preferential accumulation in the outer cells of the circle in the vicinity of the $N_{ext}$. This can be clearly appreciated in Figure 7b and c, the latter showing the dependence of the concentration of positive charge on the radial coordinate ρ (see details of this calculation in Supporting Information S9). This plot clearly shows that the positive charge of seed cations becomes concentrated near the shell (i.e., where depth profiling has found a maximum concentration of $K^+$ and other positive cations), due to the electrostatic attraction exerted by the external negative charge during the plasma treatment. It is convinient in this regard to insist on the difference of probing depth available by XPS depth profiling (i.e., 100-200 nm) and the results of calculations in Figure 7c, where the enrichment of positive ions extends to several hundred microns depth. This thickness is of the same order of magnitude as the $K^+$ enrichment zone determined by EDX analysis (c.f. Figure 3), thus supporting the validity of our Monte Carlo simulations, at least at a qualitative level.

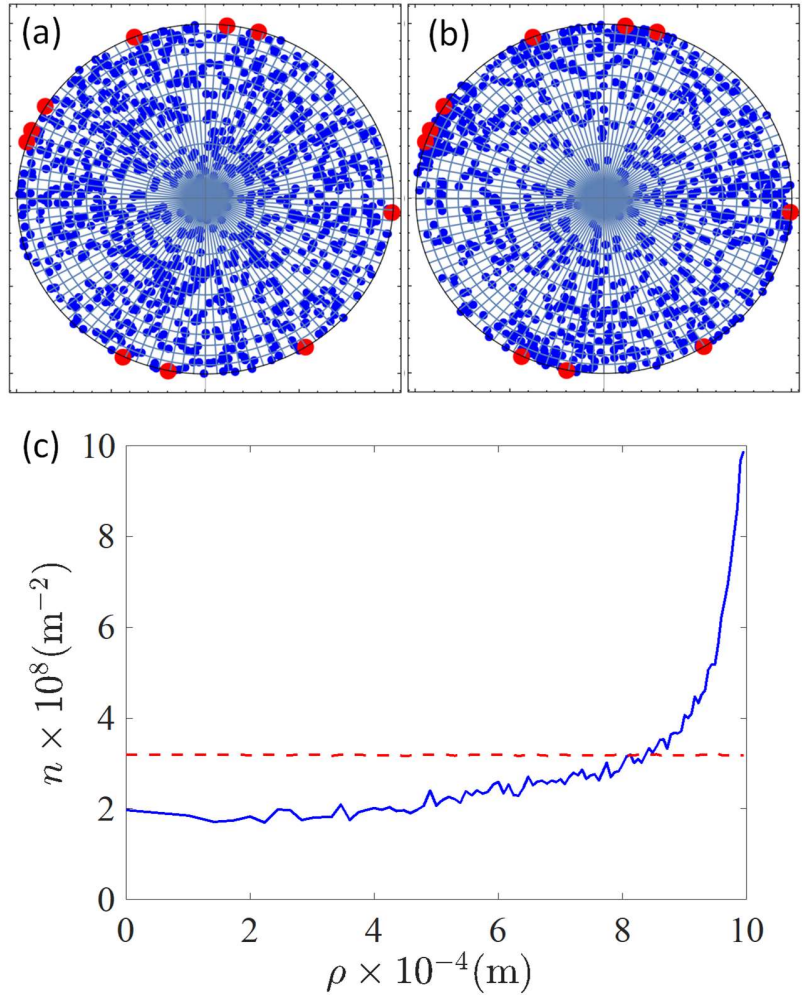

Figure 7. Typical configurations of the ions in the initial (a) and final (b) stages of the Monte Carlo simulation. Specifically, the mobile positive ions (blue) and the quenched external negative charges (red) are plotted, at the beginning (a) and the end (b) of the iteration process. Initially, the positive ions are randomly distributed in the circle, whereas the external negative charge accumulates at random locations of the shell. In the final state, an extra concentration of positive ions appears in the cells close to the negative charges. (c) Concentration of the positive charge of cations as a function of the distance to the center of the circle ρ in the equilibrium state. The solid horizontal straight line represents the theorical concentration of a uniformly distributed set of positive ions (i.e., in the initial state).

The evaluation of the variation of the system energy (i.e., U) from the initial state (c.f. Figure 7a) to the final state (c.f. Figure 7b) rendered a change from approximately $-1.25 \times 10^{-20}$ J/ion to ca. $-1.45 \times 10^{-20}$ J/ion. Convergence was reached after $6 \times 10^5$ iterations (the evolution of the computed energy of the system along the iterative convergence process can be seen in Supporting Information S9).

It is noteworthy that simulation results similar to the ones presented here were obtained substituting the homogeneous distribution of negative charge density $\sigma$ within the circle by $N$ negative point charges distributed uniformly along the cells (i.e., mimicking the distribution of positive charges in the initial state). This supports the validity of the model assumptions regarding the compensation of positive charge in the interior of the seeds. It must be also stressed that the actual values of the parameters used for the calculations (i.e., values of $N$, $N_{ext}$, $Q_{ext}$, …) affected the degree of the observed effects, but not their general, qualitative, trend. For example, the incorporation of less negative charge on the surface gave rise to a lesser accumulation of positive ions at the outer layers of the disk. Similarly, a smaller potential barrier between cells made that the equilibrium state was reached after fewer iterations. However, the pervasive common result was that the external negative charge incorporated on the surface during plasma treatment acts as a thrust to redistribute the positive ions around them, i.e., at the outer layers of the system as observed by XPS and EDX in the experiments.

### 3.5. Analysis of water ion release processes

The surface analysis of the seed hull before and after plasma treatment has revealed the occurrence of significant chemical and ion redistribution changes in a thickness of at least 200 nm (the actual thickness is not accessible by the GCIB-XPS experiment but can be estimated to be a few microns as suggested by the EDX potassium maps in Figure 3). The experiments in this section aim at determining whether the degree of ions release into water is affected by the plasma treatment or not. Experiments have monitored the release of ions from the seeds when they are immersed in pure water. Figure 8 shows the amount of cations such as $K^+$, $Na^+$, $NH_4^+$ and $Mg^+$ determined by ICP-MS in distilled water after the immersion of pristine and plasma treated seeds for 10 and 120 min (see the specific conditions in the Materials and Methods section). It is apparent that, except for $NH_4^+$, the release of cations, particularly $K^+$ and $Na^+$, is always higher from the plasma treated than from the control seeds. This result suggests that the protection and isolation functions of the hull membrane have been somehow weakened for the plasma treated seeds. We can also postulate that the enrichment of cations at the outer layers of the seed as an effect of the plasma treatment (c.f., Figures 4-6) makes more favorable their release to the water upon immersion (66, 67).

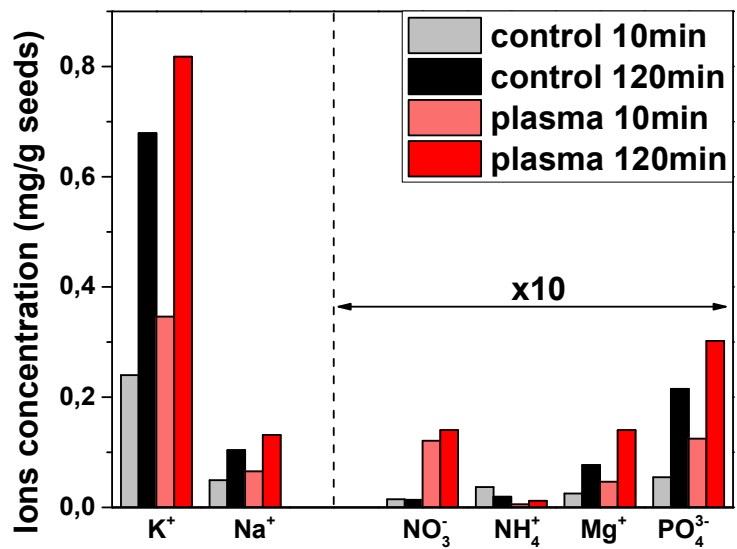

Figure 8. ICP-MS analysis of ions (mg/g) released from control and plasma treated seeds immersed in distilled water for 10 and 120 min. The concentrations of $NO_3^-$, $NH_4^+$, $Mg^{2+}$ and $PO_4^{3-}$ ions are multiplied by a factor of 10 to make them observable.

Interestingly, the very little lixiviation found for anions, particularly $PO_4^{3-}$, is also enhanced for the plasma treated seeds. The case of $NO_3^-$ is particularly relevant in this regard. Firstly, because practically no $NO_3^-$ ions were lixiviated for the control seeds (this agrees with that a little or negligible amount of free $NO_3^-$ exist in the interior of barley seeds (68)), while a net amount was detected in the plasma treated seeds. Secondly, because in the plasma treated seed the amount of $NO_3^-$ was rather similar after 10 or 120 min of seed immersion in water. The detection of lixiviated $NO_3^-$ from the plasma treated seeds is somehow contradictory with the fact that no $NO_x$ species were detected by XPS on the surface of plasma treated seeds (c.f. Figure 6), while other forms of nitrogen significantly increased with respect to the control seeds. We hypothesize that the detected nitrate forms through a chemical reaction involving the extra accumulated nitrogen in the outer layer of plasma treated seeds and the ROS species formed and adsorbed on the seeds during the air plasma treatment. The process could be described by the following reaction schematics, which would take place in the presence of water and lead to the formation of $NO_3^-$ and its release to the water where the seeds are soaked:

C-N(H) + ROS* → $NO_3^-$ ($H_2O$) (10)

The results disclosed in this work have identified some of the physical effects that may take place when seeds are exposed to plasma. Unlike most contributions in this area where focus is placed on chemical, biochemical or even gene expression effects of the plasma treatment (1, 3, 5), the evidence gathered in this investigation has shown that physical interactions are not negligible and
34

may induce the re-distribution of ions in the interior of seeds. Whether such effects can be linked with the known chemical and biochemical factors affecting the seed and young plant metabolism is a question to be addressed in specific works on this topic. Interestingly, the current paradigm to account for these biochemical or gene expression effects relies on the effect of ROS and/or RONS in triggering specific metabolic reactions contributing to increase the germination rate of seeds or to improve the seedling process (11, 43, 63). We hypothesize that the detection of $NO_3^-$ anions in the water after soaking the plasma treated seeds is a hint of a water promoted reaction between adsorbed ROS and the nitrogen species incorporated in the outer layers of seeds. Nitrate species are known as precursors of the formation of amino-acid in plants through its enzymatic reduction to nitrite and ammonia species. Also, as a signaling species responsible for reducing the dormancy of seeds and triggering the germination processes (69, 70). In this regard, the formation of $NO_3^-$ through reaction (10) would be an adjuvant factor of the germination. The redistribution of $K^+$ and other cations inside the seed might be another one. Different works have reported that exogen $K^+$ may contribute to improve the seed germination and the seedling process (71, 72). However, at present, the implications of these ion-linked processes for the control of seed metabolism and/or for triggering specific biochemical growing factors are questions open for debate that require of specific new works directly addressing these open questions.

4. **Conclusion**

The results in this work have clearly established that exposing seeds to cold atmospheric air plasma produces changes in the distribution of charged species in their interior. Specifically, it has been proved that positive ions diffuse and tend to accumulate over the outer layers of seeds in a thickness of at least 200 nm that likely extends up to some microns. The driving force for this redistribution has been associated to the incorporation onto the seed surface of negative charge, a process naturally occurring whenever a body is exposed to plasma and a plasma sheath forms around it. The simulation of the cation diffusion process with a Monte Carlo model has given support to the experimental results and has opened the way of using simple physical models to describe complex mechanisms occurring in living structures. In addition to a redistribution of positive charge, the XPS analysis of the composition of the outer layer of the seeds exposed to the plasma has proved the occurrence of important chemical modifications that extend to thicknesses of, at least, several hundred nanometers. As a hypothesis, we propose that in the presence of water, the extra nitrogen incorporated in these outer layers in the form of C-N or C-NH$_x$ species might react with ROS species to yield NO$_3^-$, a known signaling molecule known for triggering germination and reducing dormancy in seeds. This study about the redistribution of ions in plasma treated seeds might be linked with other changes in metabolic functions contributing to better understanding, control, and improvement of the seedling (germination) processes.

ACKNOWLEDGMENTS

The authors thank the projects PID2022-143120OB-I00, PID2020-114270RA-I00, PID2020-112620GB-I00 and PID2023-147916NA-I00, TED2021-130124A-I00 and TED2021-130916B-I00 funded by MCIN/AEI/10.13039/ 501100011033 and by "ERDF (FEDER)" A way of making Europe, Fondos NextgenerationEU and Plan de Recuperación, Transformación y Resiliencia". CLS thanks the University of Seville through the VI PPIT-US and "Ramon y Cajal" program funded by MCIN/AEI/10.13039/501100011033. The project leading to this article has received funding from the Consejería de Economía, Conocimiento, Empresas y Universidad de la Junta de Andalucía through the project US-1381045 as well as the EU through cohesion fund and FEDER 2014–2020 programs for financial support. Natalia Ruiz Pino acknowledges support from the FPU programme through Grant FPU2021/01764.

# Ion mobility and segregation in seed surfaces subjected to cold plasma treatments

Alvaro Perea-Brenes,[a] Natalia Ruiz-Pino,[b] Francisco Yubero,[a] Jose Luis Garcia,[c] Agustín R. Gonzalez-Elipe,[a] Ana Gomez-Ramirez,[a,b] Antonio Prados,[b*] Carmen Lopez-Santos[a,d*]

[a] Nanotechnology on Surfaces and Plasma Laboratory, Institute of Materials Science of Seville, Consejo Superior de Investigaciones Científicas-Universidad de Sevilla, Seville 41092, Spain

[b] Física Teórica, Departamento de Física Atómica, Molecular y Nuclear, Universidad de Sevilla, Apartado de Correos 1065, Seville 41080, Spain

[c] Department of Plant Biotechnology, Institute of Natural Resources and Agrobiology of Seville, Consejo Superior de Investigaciones Científicas, Seville 41012, Spain

[d] Departamento de Física Aplicada I, Escuela Politécnica Superior, Universidad de Sevilla, Seville 41011, Spain,

*mclopez@icmse.csic.es; prados@us.es

**Supporting information S1.- Scheme for the distribution of positive charges inside the seed before the plasma treatment and effect of plasma exposure.**

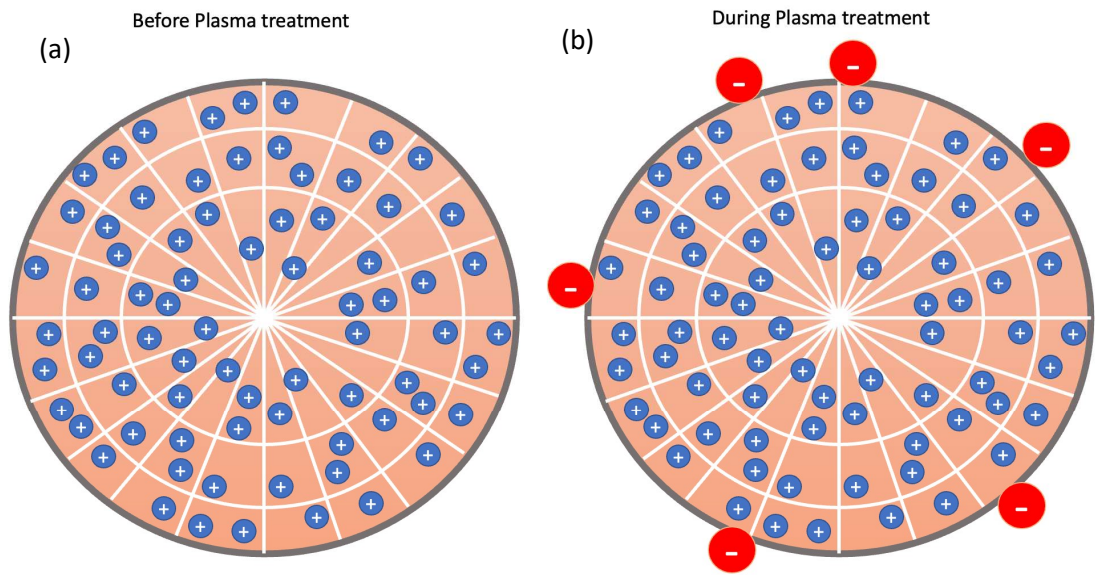

**Fig. S1:** Theoretical model of the ions distribution in the seed both before (a) and during (b) the plasma treatment. During the plasma treatment, specific locations at the seed surface become negatively charged. The configuration shown in (b) constitutes the initial state for the Monte Carlo simulation.

**Supporting information S2.- Calculation of the electric potential due to the homogeneous distribution of negative charge in the interior of seed**

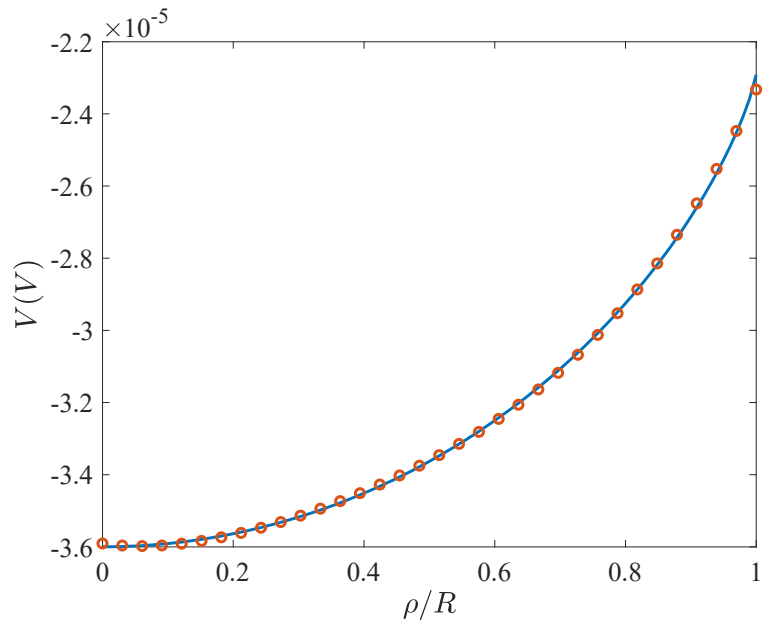

**Fig. S2: Electric potential created by the background uniformly distributed negative density of charge σ, as a function of the distance ρ to the center of the circle. The exact value is drawn in blue solid line and our estimation, employing a 4-th order polynomial function to determine V($\vec{r}$), with red circles.**

**Supporting information S3.- Monte Carlo principles enabling that the system reaches a final equilibrium state.**

The principles and procedure used to let the system in Figure S1b) to evolve proceeds according to the following principles:

We consider two different configurations of the mobile ions, $\{\vec{r}_1, \vec{r}_2, \ldots, \vec{r}_N\}$ and $\{\vec{r}_1{}', \vec{r}_2{}', \ldots, \vec{r}_N{}'\}$, and introduce $w(\{\vec{r}_1, \vec{r}_2, \ldots, \vec{r}_N\} \to \{\vec{r}_1{}', \vec{r}_2{}', \ldots, \vec{r}_N{}'\})$ as the probability for the transition from $\{\vec{r}_1, \vec{r}_2, \ldots, \vec{r}_N\}$ to $\{\vec{r}_1{}', \vec{r}_2{}', \ldots, \vec{r}_N{}'\}$ in one step. Provided that:

i) the following condition holds.

$$w(\{\vec{r}_1, \vec{r}_2, \ldots, \vec{r}_N\} \to \{\vec{r}_1{}', \vec{r}_2{}', \ldots, \vec{r}_N{}'\})e^{-\beta U(\vec{r}_1, \vec{r}_2, \ldots, \vec{r}_N)} = w(\{\vec{r}_1{}', \vec{r}_2{}', \ldots, \vec{r}_N{}'\} \to \{\vec{r}_1, \vec{r}_2, \ldots, \vec{r}_N\})e^{-\beta U(\vec{r}_1{}', \vec{r}_2{}', \ldots, \vec{r}_N{}')}, \quad (S1)$$

which is known as detailed balance, and

ii) two arbitrary configurations can be connected to a certain chain of transitions, which is known as ergodicity, the equilibrium distribution is reached in the long-time limit, i.e. after a large enough number of steps [R1]. This is the basis of the Monte Carlo chain methods that are extensively used in the simulation of mesoscopic systems [R2].

Here, we implement an effective Monte Carlo dynamics for our system, specifically a variant of the so-called Metropolis algorithm. In each step of the dynamics, a mobile positive ion, the $i$-th one, is randomly chosen from the set of $N$ ions, together with a random displacement $\vec{a}$ thereof. The following transition, from the "old" to the "new" configuration, is attempted:

$$\{\vec{r}_1, \ldots, \vec{r}_i, \ldots, \vec{r}_N\} \to \{\vec{r}_1, \ldots, \vec{r}_i + \vec{a}, \ldots, \vec{r}_N\} \quad (S2)$$

The change of internal energy for the attempted transition is thus

$$\Delta U \equiv U_{new} - U_{old} = U(\vec{r}_1, \ldots, \vec{r}_i + \vec{a}, \ldots, \vec{r}_N) - U(\vec{r}_1, \ldots, \vec{r}_i, \ldots, \vec{r}_N). \quad (S3)$$

The typical *Metropolis* algorithm assigns the attempted transition a probability $p = \min\{1, e^{-\beta \Delta U}\}$, i.e. the transition is always accepted if it decreases $U$, whereas it is accepted with probability $e^{-\beta \Delta U} < 1$ if it increases $U$. This *Metropolis* rule can be formulated according to the relation $w(\{\vec{r}_1, \ldots, \vec{r}_i, \ldots, \vec{r}_N\} \to \{\vec{r}_1, \ldots, \vec{r}_i + \vec{a}, \ldots, \vec{r}_N\}) = \min\{1, e^{-\beta \Delta U}\}$, which verifies the balance condition in Eq. S1, and thus drives the system to equilibrium. Still, in order to be more realistic, we also introduce an energy cost $E_0$ if the positive ion moves to an adjacent cell as a consequence of the attempted transition, i.e., when $\vec{r}_i \to \vec{r}_i + \vec{a}$. This accounts for an energy barrier hindering the motion of positive ions to a different cell. Therefore, we consider that

$$w(\{\vec{r_1}, \ldots, \vec{r_i}, \ldots, \vec{r_N}\} \to \{\vec{r_1}, \ldots, \vec{r_i} + \vec{a}, \ldots, \vec{r_N}\}) = e^{-\beta E_0 \delta(a,i)} \min\{1, e^{-\beta \Delta U}\}, \quad (S4)$$

in which $\delta(a,i) = 1$ if $\vec{r}_i$ and $\vec{r}_i + \vec{a}$ belong to different cells, whereas $\delta(a,i) = 0$ if $\vec{r}_i$ and $\vec{r}_i + \vec{a}$ belong to the same cell. As compared with the case $E_0 = 0$, the transitions that involve crossing a cell boundary are slowed down by the Arrhenius-like factor $e^{-\beta E_0}$. Since the reverse transition also involves crossing a cell boundary, the balance condition in Eq. S1 is preserved and this more realistic dynamics also drives the system to equilibrium.

Using our Monte Carlo algorithm, the system transits through different ion configurations, jumping from one to another with the probability just described above. In this way, we construct a *Markov* chain with detailed balance, and thus the final—for very long times—state of the system corresponds to equilibrium at the bath temperature $T$, described by the canonical distribution, which minimizes the free energy. For low temperatures, the system would reach the state with minimal energy (or, at least, a local minimum), since the entropic contribution to the free energy becomes negligible. This is the equilibrium state resulting from the accumulation of extra negative charge on the surface of seeds during the plasma treatment.

[R1] V. Kampen, N. Godfried. Stochastic Processes In Physics and Chemistry. 3rd ed. Amsterdam: Elsevier, 2007 ISBN 9780080475363.

[R2] D. Frenkel, B. Smit. Understanding Molecular Simulation: From Algorithms to Applications. 3rd ed. Elsevier, 2023 ISBN 9780323902922.

**Supporting information S4.- SEM and EDX analysis of barley seeds.**

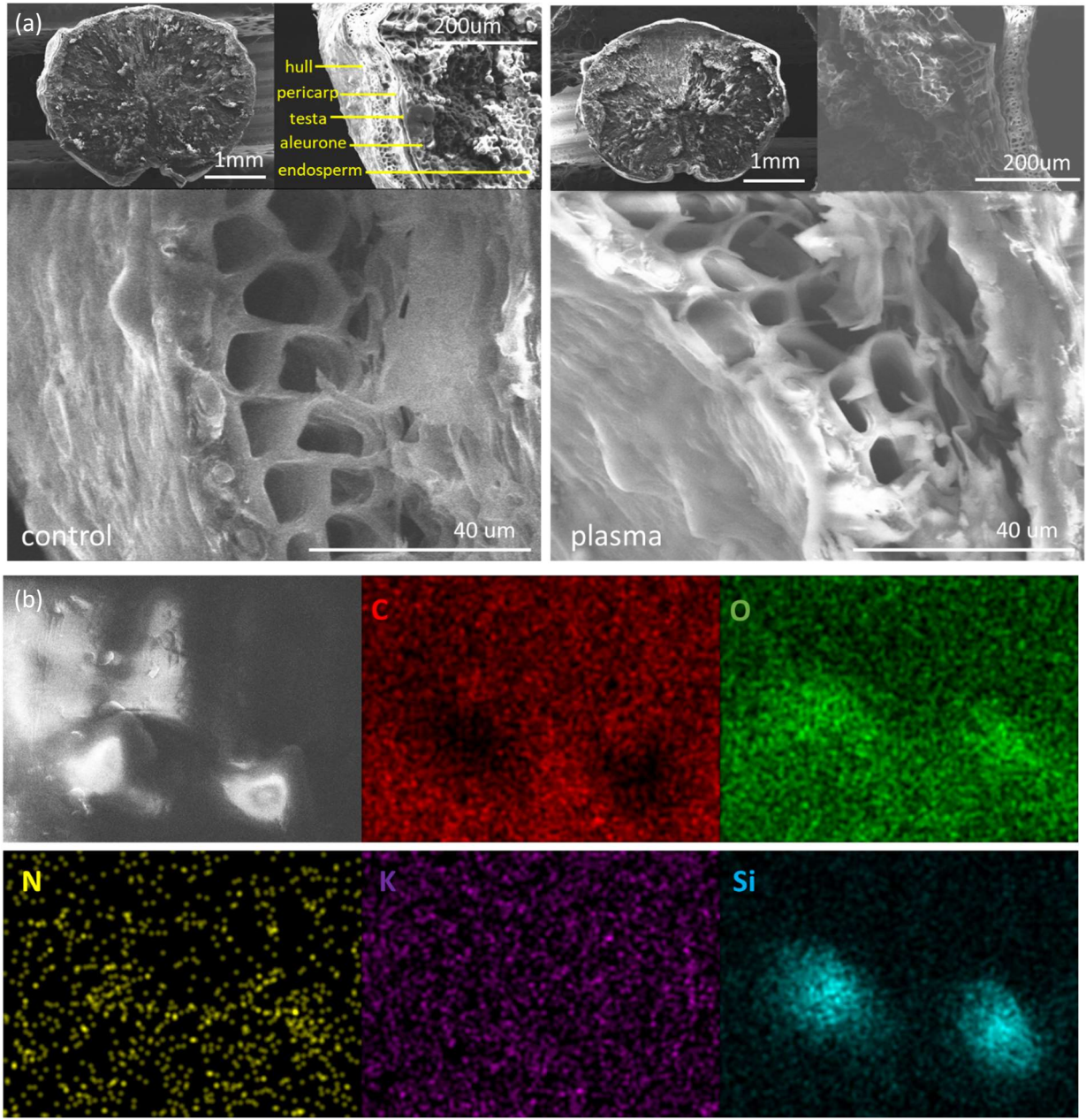

**Fig. S4.** (a) SEM images of cuts of the barley seeds highlighting the internal structure of the pericarp, indicating the main identified layers (yellow color). No significant differences in morphology for length scales of several microns are observed between the control and the plasma treated seeds. (b) SEM image and EDX analysis of the surface of the seed in the form of C, O, N, K and Si maps, corresponding to a barley seed after a plasma treatment for 3min. The presence of $SiO_2$ agglomerates on the barley seed surface is clearly appreciated in these images

**Supporting information S5.- Additional EDX analysis of barley seeds cuts**

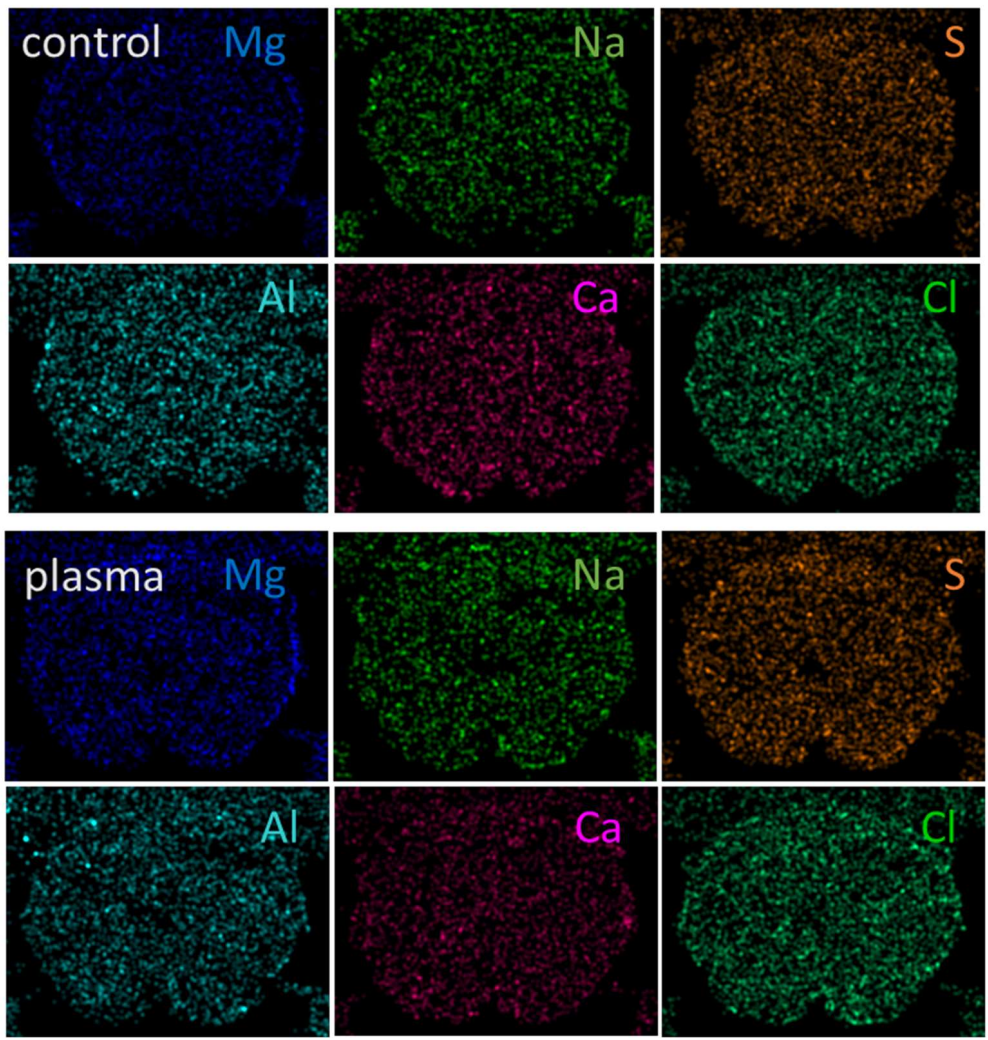

Fig. S5. EDX additional maps of the seed cuts before and after the plasma treatment corresponding to the minority elements detected: Mg, Na, S, Al, Ca, and Cl.

**Supporting information S6.- Etching profiles according to protocol 2 for barley seeds before and after plasma treatment.**

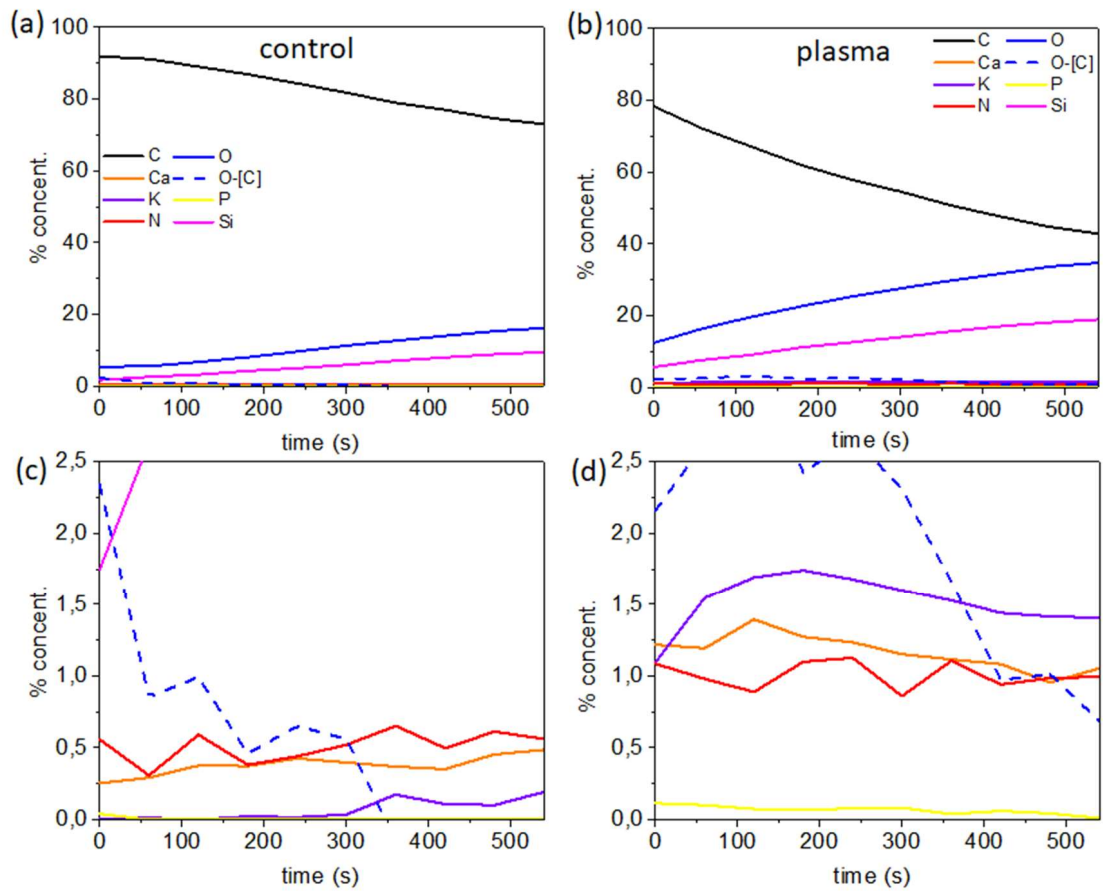

Fig. S6. Element depth profiles according to protocol 2 for the reference (a, c) and plasma treated (b,d) seeds. Dashed lines represent the depth profile of oxygen associated to the organic component of the outer layers of seeds (i.e., once discounting the oxygen associated to silicon). These depth profiles confirm that the enrichment in N, P and K after the plasma treatment extends to a depth of at least 200 nm.

**Supporting information S7.- XPS spectra of minority elements recorded after GCIB depth profiling of pristine and plasma treated barley seeds.**

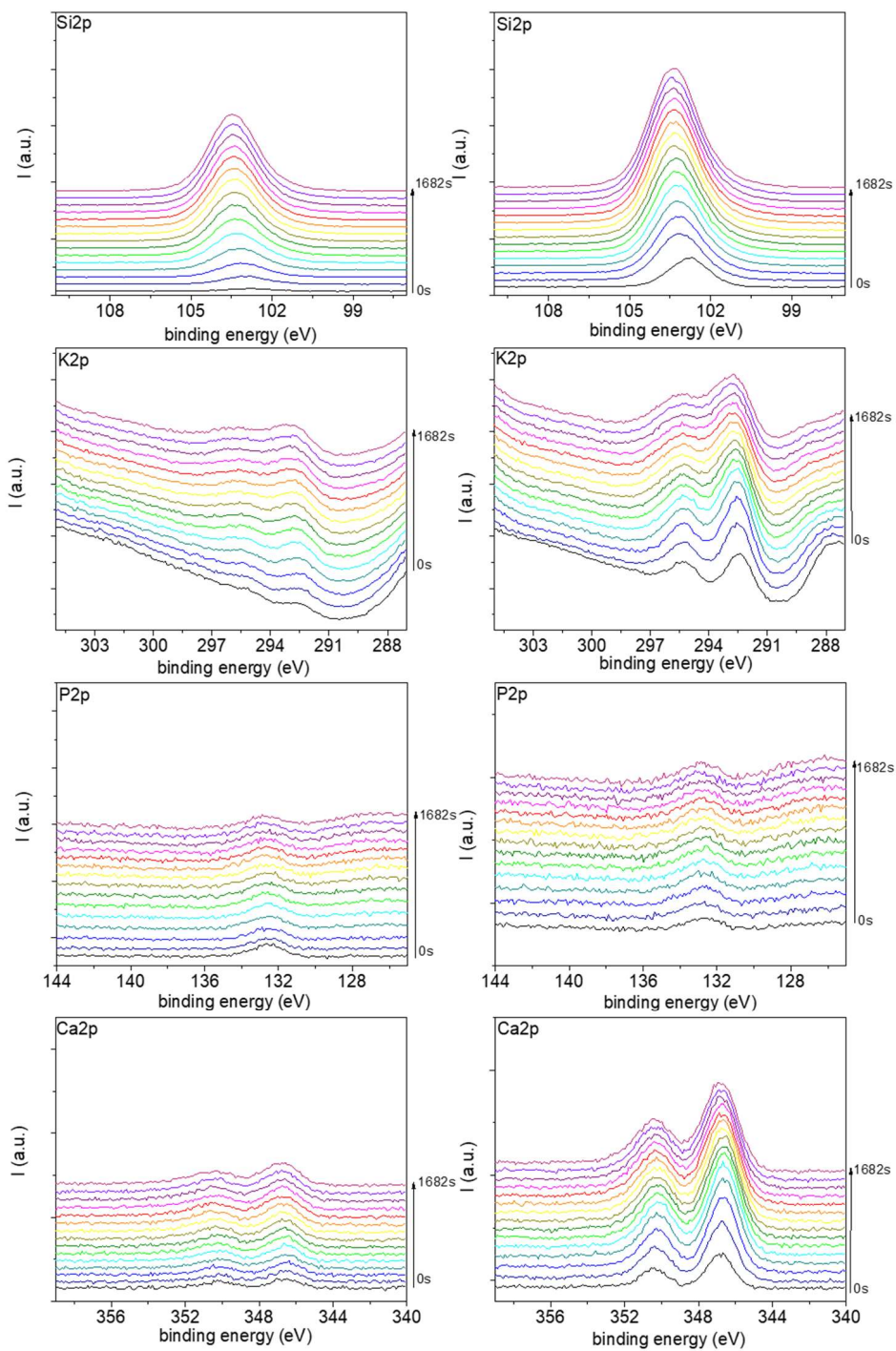

**Fig. S7. Selected Si 2p, K 2p, P 2p and Ca 2p spectra taken after GCIB depth profiling (according to the etching time indicated with the arrow) for control (a) and plasma treated (b) barley seeds.**

**Supporting information S8.- Summary of the binding energies and atomic concentration of the functional groups at the barley surface before and after the plasma treatment obtained by fitting the XPS spectra**

| XPS functional group | Binding energy (eV) | Fitted At. Concentration (%) | |
|:---:|:---:|:---:|:---:|
| | | Untreated | Plasma treated |
| -C-C | 284.5 | 66.1 | 56.1 |
| -C-O | 286.0 | 27.8 | 33.3 |
| -C=O | 287.5 | 9.1 | 10.6 |
| C-O | 532.7 | 11.4 | 91.8 |
| C=O | 533.2 | 88.1 | 7.7 |
| N-O | 536.2 | 0.5 | 0.5 |
| Si-C-O | 102.8 | 100 | 73.4 |
| Si-O | 103.4 | 0 | 26.6 |
| K-Cl/-SO$_4$ | 292.8 | 100 | 100 |
| Ca-O | 346.7 | 100 | 100 |
| P-O | 132.7 | 100 | 100 |

**Supporting information S9.- Evolution of the internal energy per ion.**

Fig. S9 showcases the evolution of the internal energy $U/N$ per (positive mobile) ion $U$ as a function of the number of algorithm iterations. This plot is not a time evolution, but a representation of the Monte Carlo dynamics that, following the energy premises in eqs. 3-7 in the main text is designed to drive the system towards equilibrium. The plot shows that, on average, the energy decreases from the value in the initial state configuration to the final state, becoming roughly constant after 6x10⁵ steps (although there are thermal fluctuations in the final state, these fluctuations do not alter the basic tendency revealed by the Monte Carlo dynamics).

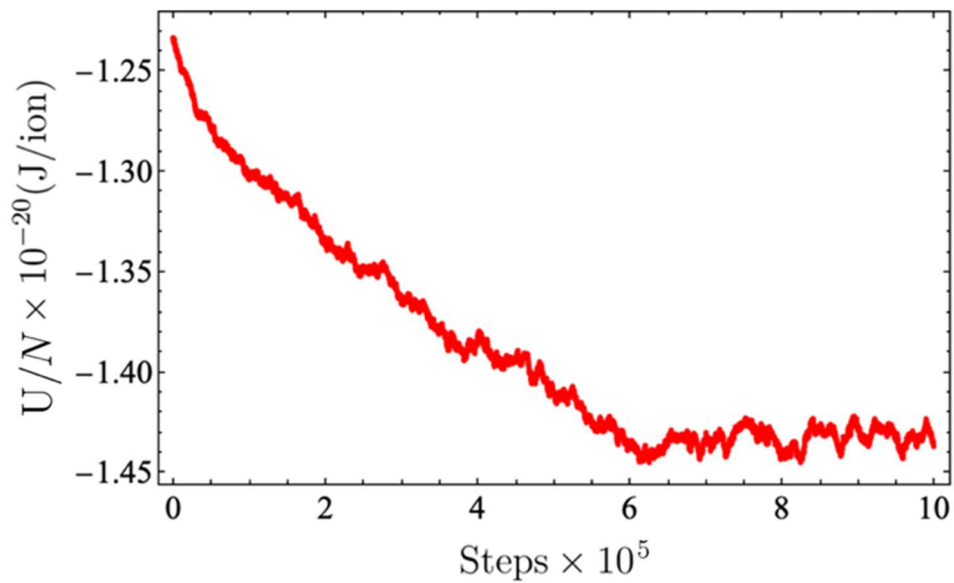

**Fig. S9. Evolution of the internal energy of the positive ions in the Monte Carlo dynamics. The internal energy is plotted as a function of the number of steps in the algorithm. The energy decreases with the number of steps, until it reaches, on average, a constant value for a number of steps $> 6 \times 10^5$, corresponding to the equilibrium state.**

**Supporting information S10.- Calculation of the distribution of positive charge in the final state of the system.**

The following procedure was used to evaluate the distribution of positive charge in the final state (c.f., Figure 7b in the main text) as a function of the radial coordinate $\rho$. Taking the numerical density $n$ as the number of ions per unit area, $N(\rho, \rho + \Delta\rho)$ will represent the number of mobile positive ions in an annulus of radii $\rho$ and $\rho + \Delta\rho$. Then

$$N(\rho, \rho + \Delta\rho) = n(\rho)\pi[(\rho + \Delta\rho)^2 - \rho^2] \simeq n(\rho)2\pi\rho\Delta\rho \quad (S5)$$

To improve the statistics for $n(\rho)$, it has been averaged over all configurations of positive ion distributions calculated from Step $= 6 \times 10^5$ to Step $= 10^6$, i.e., once the equilibrium state has been reached.